\title{Generalized Concatenated Types of Codes for
Erasure Correction}
\author{Mario Blaum and Steven Hetzler\\
IBM Almaden Research Center\\
San Jose, CA 95120
}
\documentclass[12pt]{article}
\usepackage{latexsym}
\usepackage{multirow}
\usepackage{amsmath}
 \newtheorem{theo}{Theorem}[section]

 \newtheorem{defin}{Definition}[section]
 \newtheorem{ex}{Example}[section]
 \newtheorem{alg}{Algorithm}[section]
 \newtheorem{cor}{Corollary}[section]
 
\newtheorem{COROLLARY}{\indent Corollary}

\newtheorem{EXAMPLE}{\indent Example}

\newtheorem{THEOREM}{\indent Theorem}

\newtheorem{REMARK}{\indent Remark}

\newcommand{\ga}{\mbox{$\gamma$}}

\newcommand{\fullstop}{\hspace{-0.85em} {\bf .}}

\newcommand{\uv}{\mbox{$\underline{v}$}}

\newcommand{\us}{\mbox{$\underline{s}$}}
\newcommand{\uu}{\mbox{$\underline{u}$}}

\newcommand{\hH}{\mbox{$\hat{H}$}}
\newcommand{\hs}{\mbox{$\hat{s}$}}

\newcommand{\ra}{\mbox{$\rightarrow$}}

\newcommand{\al}{\mbox{$\alpha$}}
\newcommand{\si}{\mbox{$\sigma$}}
\newcommand{\eq}{\mbox{$\, =\,$}}

\newcommand{\tr}{\mbox{$\triangle$}}
\newcommand{\qed}{\hfill$\Box$\\[1ex]}
\newcommand{\pf}{{\bf Proof: }}

\newcommand{\uw}{\mbox{$\underline{w}$}}
\newcommand{\uzero}{\mbox{$\underline{0}$}}

\newcommand{\xor}{\mbox{$\,\oplus\,$}}

\newcommand{\C}{\mbox{${\cal C}$}}

\newcommand{\tH}{\mbox{$\tilde{H}$}}

\newcommand{\cO}{\mbox{${\cal O}$}}

\newcommand{\br}{\\ }
\newcommand{\ce}{\begin{center}}
\newcommand{\cen}{\end{center}}

\newcommand{\ipb}{\begin{description}}
\newcommand{\ipn}{\end{description}}

\newcommand{\qb}{\begin{quote}}
\newcommand{\qn}{\end{quote}}

\newcommand{\tp}{\begin{titlepage}}
\newcommand{\tpn}{\end{titlepage}}
\newcommand{\zb}{\begin{figure}[hbtp]}
\newcommand{\zn}{\end{figure}}

\newcommand{\EQX}[1]{\begin{equation}\label{#1}}
\newcommand{\ENX}{\end{equation}}
\newcommand{\EQL}{\begin{eqnarray*}}
\newcommand{\ENL}{\end{eqnarray*}}
\newcommand{\EQLX}[1]{\begin{eqnarray}\label{#1}}
\newcommand{\ENLX}{\end{eqnarray}}

\newcommand{\open}{\begin{document}}
\newcommand{\close}{\end{document}}
\newcommand{\lfcr}[1]{\br\hspace*{#1em}}
\newenvironment{mat}[1]
{\left[\begin{array}{#1}}{\end{array}\right]}
\newcommand{\GAMMA}{\Gamma}
\newcommand{\DELTA}{\Delta}
\newcommand{\THETA}{\Theta}
\newcommand{\LAMBDA}{\Lambda}
\newcommand{\XI}{\Xi}
\newcommand{\PI}{\Pi}
\newcommand{\SIGMA}{\Sigma}
\newcommand{\UPSILON}{\Upsilon}
\newcommand{\PHI}{\Phi}
\newcommand{\PSI}{\Psi}
\newcommand{\OMEGA}{\Omega}
\newcommand{\bldgreek}[1]{\mbox{\boldmath $#1$}}
\newcommand{\bldbeta}{\bldgreek{\beta}}
\newcommand{\bldgamma}{\bldgreek{\gamma}}
\newcommand{\blddelta}{\bldgreek{\delta}}
\newcommand{\bldepsilon}{\bldgreek{\epsilon}}
\newcommand{\bldvarepsilon}{\bldgreek{\varepsilon}}
\newcommand{\bldzeta}{\bldgreek{\zeta}}
\newcommand{\bldeta}{\bldgreek{\eta}}
\newcommand{\bldtheta}{\bldgreek{\theta}}
\newcommand{\bldvartheta}{\bldgreek{\vartheta}}
\newcommand{\bldiota}{\bldgreek{\iota}}
\newcommand{\bldkappa}{\bldgreek{\kappa}}
\newcommand{\bldlambda}{\bldgreek{\lambda}}
\newcommand{\bldmu}{\bldgreek{\mu}}
\newcommand{\bldnu}{\bldgreek{\nu}}
\newcommand{\bldxi}{\bldgreek{\xi}}
\newcommand{\bldpi}{\bldgreek{\pi}}
\newcommand{\bldvarpi}{\bldgreek{\varpi}}
\newcommand{\bldrho}{\bldgreek{\rho}}
\newcommand{\bldvarrho}{\bldgreek{\varrho}}
\newcommand{\bldsigma}{\bldgreek{\sigma}}
\newcommand{\bldvarsigma}{\bldgreek{\varsigma}}
\newcommand{\bldtau}{\bldgreek{\tau}}
\newcommand{\bldupsilon}{\bldgreek{\upsilon}}
\newcommand{\bldphi}{\bldgreek{\phi}}
\newcommand{\bldvarphi}{\bldgreek{\varphi}}
\newcommand{\bldchi}{\bldgreek{\chi}}
\newcommand{\bldpsi}{\bldgreek{\psi}}
\newcommand{\bldomega}{\bldgreek{\omega}}
\begin{document}
\parindent=10pt
\maketitle
\begin{abstract}
Generalized Concatenated (GC), also known as Integrated Interleaved
(II) Codes, are studied from an erasure correction point of view making them
useful for Redundant Arrays of Independent Disks (RAID) types of
architectures combining global and local
properties. The fundamental erasure-correcting properties of
the codes are proven and
efficient encoding and decoding algorithms are provided. Although less powerful
than the recently developed PMDS codes, this implementation has the
advantage of allowing generalization to any range of parameters while
the size of the field is much smaller than the one required for PMDS
codes.

\vspace{.3cm}

\end{abstract}

\section{Introduction}
\label{Introduction}

Considerable interest has arisen lately in coding schemes that combine
local and global properties. Applications like Redundant Arrays of
Independent Disks (RAID)
architectures~\cite{bhh}\cite{hsx}\cite{ll}\cite {pb} are an example
of this interest.
In effect, given an array of disks, a regular RAID architecture like,
say, RAID 5, protects against a total disk (or, more in
general, a storage device) failure. This is
simply done by XORing the
data devices in order to obtain a parity device (in this
paper, we do not distinguish between RAID 4 and RAID 5, since this
distinction is
not relevant to our discussion). Then, if a storage
device fails, its contents can be recovered by XORing the
surviving devices.

A problem with this approach is that there may be individual sectors in the surviving
devices that have failed due to uncorrectable bit errors (what is known as
silent failures), a problem with Solid State Devices (SSDs), that
deteriorate as a function of time and of usage. 
In that case, one
individual sector that has failed will cause data loss in the
presence of a total device failure.

A method around this situation is using RAID 6: adding a
second parity device allows for correction of most individual sector
failures in the presence of a total device failure. The
drawback of this approach is that it is wasteful: if for example a
few extra sectors need to be recovered in addition to all the sectors
corresponding to the failed device, it is desirable to optimize the
redundancy necessary for doing so.

Codes dealing with this problem are the Partial MDS (PMDS)
codes~\cite{b}\cite{bhh}\cite{bp}\cite{bpsy}\cite{ghjy}\cite{hsx}
(in~\cite{ghjy}\cite{hsx}, PMDS
codes are called Maximally Recoverable codes), sector-disk (SD)
codes \cite{pbh}\cite{pb}, Locally Recoverable Codes (LRC)~\cite{sa}
and STAIR codes~\cite{ll}. 

In general, we consider an $m\times n$ array. The parameter $n$
represents the number of devices and $m$ represents the size of a
stripe: $m$ is repeated a number of times throughout the array and
each $m\times n$ stripe is decoded independently of the others.

The codes to be described in this paper are weaker than those
in~\cite{bhh}\cite{ghjy}\cite{hsx}, in the sense that there are some
erasure patterns that they cannot correct for the same amount of
redundancy.
However, they can be generalized to any set of parameters and, more
importantly, they are
simpler to implement, since they require a finite field
${\rm GF}(2^b)$ of size $2^b> n$, the length of the rows, while the
codes in~\cite{hsx} require size $2^b> mn$, the total length of the
array (and the known constructions require much
larger fields~\cite{bhh}\cite{ghjy}\cite{hsx}). Similar considerations
inspired the recent STAIR codes~\cite{ll}. In~\cite{rv}, different
combinations of local and 
global failures, involving either erasures and errors, are corrected
using probabilistic methods by exploiting the rank of the error
arrays. In~\cite{sa}, the data is encoded using
a global RS code, and it is divided into parity groups that are
independently encoded from the RS symbols. The Zigzag
codes~\cite{twb} keep the MDS property and optimize the minimum
number of updates in the presence of one failure, but the parameter
$m$ is exponential on the number of devices $n$. In~\cite{cs}, a new
probabilistic method is studied for decoding arrays using
two-dimensional LDPC codes.

In order to illustrate our discussion, consider a (1,2) PMDS code
over $4\times 5$ 
arrays~\cite{b}. The code can correct an erasure in each row, and in
addition two extra erasures anywhere. Below are two examples of
erasure-patterns that can be corrected, where the erasures are
indicated by $X$:

$$
\begin{array}{cc}
\begin{array}{|c|c|c|c|c|}
\hline
X&\phantom{X}&\phantom{X}&\phantom{X}&\phantom{X}\\
\hline
\phantom{X}&X&\phantom{X}&\phantom{X}&X\\
\hline
\phantom{X}&\phantom{X}&\phantom{X}&\phantom{X}&X\\
\hline
\phantom{X}&\phantom{X}&X&\phantom{X}&X\\
\hline
\end{array}
&
\begin{array}{|c|c|c|c|c|}
\hline
X&\phantom{X}&\phantom{X}&\phantom{X}&\phantom{X}\\
\hline
\phantom{X}&X&\phantom{X}&X&X\\
\hline
\phantom{X}&\phantom{X}&\phantom{X}&\phantom{X}&X\\
\hline
\phantom{X}&\phantom{X}&\phantom{X}&\phantom{X}&X\\
\hline
\end{array}
\end{array}
$$

The array on the left has two rows with two erasures each, while the
array on the right has a row with three erasures. The remaining rows
have one erasure each, that is corrected by a horizontal parity-check
code. The PMDS codes dealing with these type of errors, as presented
in~\cite{b}, require a field of size at least $2mn$ (these codes were
extended in~\cite{bpsy}). The codes to be presented will require a
field of size at least $n+1$ only, one more than the length of the
rows, but will 
correct, in this example, either the arrays on the left, or those on
the right, but not both simultaneously (or, they can correct both
simultaneously by using more redundancy). However, the codes can be
extended to any set of parameters. 

Actually, codes having the desired characteristics were created for a
different application. Those are the so called Generalized
Concatenated (GC) codes~\cite{bz}\cite{z}. GC codes were presented in
a form more suitable for implementation by the so called Integrated
Interleaved (II) codes~\cite{hapkt}\cite{tk}. Here we want to adapt an
II type of approach as an erasure-correcting code to deal with the
problem of local and global parities. Some of the uses of GC codes
for erasure-correction in RAID type of architectures were presented
in~\cite{bhh2}. The description of the codes to be presented in this
paper is based on their parity-check matrices.

In the next section we give the formal definition of the codes, we
illustrate them with several examples and then we
prove their basic property
in Theorem~\ref{theo1}. In Section~\ref{encdec} we present efficient
encoding and decoding algorithms that are based on a divide and
conquer approach: at each step an individual Reed-Solomon (RS)
code~\cite{ms}
of length $n$
is decoded for erasures, starting by the rows of the array having the
less erasures. The procedure is much faster than by solving at once
all the erasures using a linear system of equations based on the
parity-check matrix. We end the paper by drawing some conclusions.

\section{Generalized Concatenated (GC) Codes as Erasure-Correcting Codes}
\label{GC}
The GC codes that we describe in this section are $m\times n$ array
codes with symbols in a finite field ${\rm GF}(2^b)$, where $2^b>
n$. In fact, the codes can be described over any finite field of
characteristic $p$, $p$ a prime number, but we keep $p\eq 2$ for
simplicity and because it is the case more relevant in applications.
Reading the 
symbols horizontally in a row-wise manner gives a code of length
$mn$. We will describe the GC codes by providing their parity-check
matrices. We will then give the erasure-correcting capability of the
codes by referring to erasures per row. We will use interchangeably the
array and the row-wise vector structure of the code throughout the paper.


Denote by $I_m$ the $m\times m$ identity matrix and by $A\otimes B$ the
Kronecker product~\cite{w} of matrices $A$ and $B$. Next we give a
formal definition of $t$-level GC codes.

\begin{defin}
\label{defGC}
{\rm Let $m\leq n$ be integers, and $\al\in {\rm GF}(2^b)$ an element of
order $\cO(\al)\geq n$ (if $\al$ is primitive, $\cO(\al)\eq 2^b-1$).
Consider the matrices
\begin{eqnarray}
\label{hunl}
H(u,n;\ell) &=&
\left(
\begin{array}{cccccc}
\al^{(n-1)\ell}&\al^{(n-2)\ell}&\ldots &\al^{2\ell}&\al^{\ell}&1\\
\al^{(n-1)(\ell+1)}&\al^{(n-2)(\ell+1)}&\ldots &\al^{2(\ell+1)}&\al^{\ell+1}&1\\
\vdots &\vdots &\ddots &\vdots &\vdots \\
\al^{(n-1)(\ell+u-1)}&\al^{(n-2)(\ell+u-1)}&\ldots &\al^{2(\ell+u-1)}&\al^{\ell+u-1}&1\\
\end{array}
\right)
\end{eqnarray}
and
\begin{eqnarray}
\label{sml}
\hH(s,m;\ell) &=&
\left(
\begin{array}{cccccc}
1&\al^{-\ell}&\al^{-2\ell}&\ldots &\al^{-(m-2)\ell}&\al^{-(m-1)\ell}\\
1&\al^{-(\ell+1)}&\al^{-2(\ell+1)}&\ldots &\al^{-(m-2)(\ell+1)}&\al^{-(m-1)(\ell+1)}\\
\vdots &\vdots &\vdots & \ddots &\vdots &\vdots \\
1&\al^{-(\ell+s-1)}&\al^{-2(\ell+s-1)}&\ldots &\al^{-(m-2)(\ell+s-1)}&\al^{-(m-1)(\ell+s-1)}\\
\end{array}
\right).
\end{eqnarray}

Let $\uu$ be a vector of non-decreasing integers and length
$m\eq s_0+s_1+\cdots +s_{t-1}$ as follows:

\begin{eqnarray}
\label{equu}
\uu &=&
\left(\overbrace{u_0,u_0,\ldots,u_0}^{s_0},\overbrace{u_1,u_1,\ldots,u_1}^{s_1},\ldots,
\overbrace{u_{t-1},u_{t-1},\ldots,u_{t-1}}^{s_{t-1}}\right),
\end{eqnarray}
where $t\geq 1$, $s_i\geq 1$ for $0\leq i\leq t-1$ and $1\leq
u_0<u_1<\ldots <u_{t-1}\leq n-1$. Let $\hs_i\eq
\sum_{j=i}^{t-1}s_j$, $0\leq i\leq t-1$ (notice that $m\eq \hs_0$).
We say that
the $[mn,mn-\sum_{i=0}^{t-1}u_is_i]$ code $\C(n;\uu)$  whose
parity-check matrix is given by the $\left(\sum_{i=0}^{t-1}u_is_i\right)\times mn$ matrix
\begin{eqnarray}
\label{eqmain}
{\bf H} (n;\uu) &=&
\left(
\begin{array}{rcl}
I_{m}&\otimes & H(u_0,n;0)\\
\hH(s_{t-1},m;0)&\otimes & H(u_{t-1}-u_0,n;u_0)\\
\hH(s_{t-2},m;\hs_{t-1})&\otimes & H(u_{t-2}-u_0,n;u_0)\\
\hH(s_{t-3},m;\hs_{t-2})&\otimes & H(u_{t-3}-u_0,n;u_0)\\
&\vdots&\\
\hH(s_{1},m;\hs_2)&\otimes & H(u_{1}-u_0,n;u_0)\\
\end{array}
\right)
\end{eqnarray}
is a $t$-level GC code.
}
\end{defin}

It would remain to be proven that the $\sum_{i=0}^{t-1}u_is_i$ rows
of matrix ${\bf H} (n;\uu)$ are linearly independent, but this will
arise as a consequence of Theorem~\ref{theo1} to be stated below.

Although~(\ref{eqmain}) provides for a compact description of the
parity-check matrix ${\bf H} (n;\uu)$, it is not easy to
visualize. Below we give a more explicit form of~(\ref{eqmain}). Let
$H_0\eq H(u_0,n;0)$ and $H_j\eq H(u_{j}-u_0,n;u_0)$ as given
by~(\ref{hunl}) for $1\leq
j\leq t-1$. Then,
\begin{eqnarray}
\label{eqmain2}
{\bf H} (n;\uu) &=&
\left(
\begin{array}{c|c|c|c}
H_0&\uzero &\ldots &\uzero\\
\uzero&H_0&\ldots &\uzero\\
\vdots &\vdots &\ddots &\vdots\\
\uzero&\uzero&\ldots &H_0\\
\hline
H_{t-1}&H_{t-1}&\ldots
&H_{t-1}\\
H_{t-1}&\al^{-1}H_{t-1}&\ldots
&\al^{-(m-1)}H_{t-1}\\
\vdots &\vdots &\ddots &\vdots\\
H_{t-1}&\al^{-(\hs_{t-1}-1)}H_{t-1}&\ldots
&\al^{-(m-1)(\hs_{t-1}-1)}H_{t-1}\\
\hline
H_{t-2}&\al^{-\hs_{t-1}}H_{t-2}&\ldots
&\al^{-(m-1)\hs_{t-1}}H_{t-2}\\
H_{t-2}&\al^{-(\hs_{t-1}+1)}H_{t-2}&\ldots
&\al^{-(m-1)(\hs_{t-1}+1)}H_{t-2}\\
\vdots &\vdots &\ddots &\vdots\\
H_{t-2}&\al^{-(\hs_{t-2}-1)}H_{t-2}&\ldots
&\al^{-(m-1)(\hs_{t-2}-1)}H_{t-2}\\
\hline
\vdots &\vdots &\ddots &\vdots\\
\hline
H_{i}&\al^{-\hs_{i+1}}H_{i}&\ldots
&\al^{-(m-1)\hs_{i+1}}H_{i}\\
H_{i}&\al^{-(\hs_{i+1}+1)}H_{i}&\ldots
&\al^{-(m-1)(\hs_{i+1}+1)}H_{i}\\
\vdots &\vdots &\ddots &\vdots\\
H_{i}&\al^{-(\hs_i-1)}H_{i}&\ldots
&\al^{-(m-1)(\hs_i-1)}H_{i}\\
\hline
\vdots &\vdots &\ddots &\vdots\\
\hline
H_{1}&\al^{-\hs_2}H_{1}&\ldots
&\al^{-(m-1)\hs_2}H_{1}\\
H_{1}&\al^{-(\hs_2+1)}H_{1}&\ldots
&\al^{-(m-1)(\hs_2+1)}H_{1}\\
\vdots &\vdots &\ddots &\vdots\\
H_{1}&\al^{-(\hs_1-1)}H_{1}&\ldots
&\al^{-(m-1)(\hs_1-1)}H_{1}\\
\end{array}
\right)
\end{eqnarray}

Let us illustrate the construction of ${\bf H} (n;\uu)$ with some
examples.

\begin{ex}
\label{ex1}
{\em
Assume $t\eq 1$, i.e., $\uu\eq
\left(\overbrace{u_0,u_0,\ldots,u_0}^{s_0}\right)$ and
$\C(n;\uu)$ is a 1-level GC code. Then, according
to~(\ref{eqmain}) and~(\ref{eqmain2}),

\begin{eqnarray*}
{\bf H} (n;\uu) &=&
\left(
\begin{array}{c}
I_{s_0}\otimes H(u_0,n;0)\\
\end{array}
\right)\\
&=&
\left(
\begin{array}{c|c|c|c}
H_0&\uzero &\ldots &\uzero\\
\uzero&H_0&\ldots &\uzero\\
\vdots &\vdots &\ddots &\vdots\\
\uzero&\uzero&\ldots &H_0\\
\end{array}
\right).
\end{eqnarray*}
This one is a trivial case, since it corresponds to $s_0$ RS
codewords of length $n$ one after the other, each codeword having
$u_0$ parity symbols. 

\qed
}
\end{ex}

\begin{ex}
\label{ex2}
{\em
Assume $t\eq 2$, i.e., $\uu\eq
\left(\overbrace{u_0,u_0,\ldots,u_0}^{s_0},\overbrace{u_1,u_1,\ldots,u_1}^{s_1}\right)$
and
$\C(n;\uu)$ is a 2-level GC code.
Then, according to~(\ref{eqmain}) and~(\ref{eqmain2}),

\begin{eqnarray}
\nonumber
{\bf H} (n;\uu) &=&
\left(
\begin{array}{rcl}
I_{s_0+s_1}&\otimes &H(u_0,n;0)\\
\hH(s_{1},m;0)&\otimes & H(u_{1}-u_0,n;u_0)\\
\end{array}
\right)\\
\label{level2}
&=&
\left(
\begin{array}{c|c|c|c}
H_0&\uzero &\ldots &\uzero\\
\uzero&H_0&\ldots &\uzero\\
\vdots &\vdots &\ddots &\vdots\\
\uzero&\uzero&\ldots &H_0\\
\hline
H_{1}&H_{1}&\ldots
&H_{1}\\
H_{1}&\al^{-1}H_{1}&\ldots
&\al^{-(m-1)}H_{1}\\
H_{1}&\al^{-2}H_{1}&\ldots
&\al^{-2(m-1)}H_{1}\\
\vdots &\vdots &\ddots &\vdots\\
H_{1}&\al^{-(s_{1}-1)}H_{1}&\ldots
&\al^{-(m-1)(s_{1}-1)}H_{1}\\
\end{array}
\right)
\end{eqnarray}
The parity-check matrix of a 2-level GC code was also presented in~\cite{hl}.

Let us take now some concrete examples of a 2-level GC code. Take
$\uu\eq (1,1,3,3)$, i.e., $u_0\eq 1$, $u_1\eq 3$, $s_0\eq s_1\eq 2$.
Then, according to~(\ref{level2}), the
parity-check matrix ${\bf H} (5;(1,1,3,3))$
of the 2-level code
$\C(5;(1,1,3,3))$ is given by

\begin{eqnarray*}
{\bf H} (5;(1,1,3,3))&=&
\left(
\begin{array}{rcl}
I_{4}&\otimes & H(1,5;0)\\
\hH(2,4;0)&\otimes & H(2,5;1)\\
\end{array}
\right).
\end{eqnarray*}

Notice that

\begin{eqnarray*}
H(1,5;0)\;\;\eq\;\; H_0&=&
\left(
\begin{array}{ccccc}
1&1&1&1&1\\
\end{array}
\right),
\end{eqnarray*}
\begin{eqnarray*}
H(2,5;1)\;\;\eq\;\; H_1&=&
\left(
\begin{array}{ccccc}
\al^4&\al^3&\al^2&\al&1\\
\al^8&\al^6&\al^4&\al^2&1\\
\end{array}
\right)
\end{eqnarray*}
and
\begin{eqnarray*}
\hH(2,4;0)&=&
\left(
\begin{array}{cccc}
1&1&1&1\\
1&\al^{-1}&\al^{-2}&\al^{-3}\\
\end{array}
\right).
\end{eqnarray*}

Explicitly, according to~(\ref{level2}),

\begin{eqnarray*}
{\bf H} (5;(1,1,3,3))&=&
\left(
\begin{array}{c|c|c|c}
H_0 &\uzero &\uzero &\uzero \\
\uzero &H_0 &\uzero &\uzero \\
\uzero &\uzero &H_0 &\uzero \\
\uzero &\uzero &\uzero &H_0 \\
\hline
H_1&H_1&H_1&H_1\\
H_1&\al^{-1}H_1&\al^{-2}H_1&\al^{-3}H_1\\
\end{array}
\right),
\end{eqnarray*}
thus, ${\bf H} (5;(1,1,3,3))$ is the matrix
\begin{eqnarray*}
\left(
\begin{array}{ccccc|ccccc|ccccc|ccccc}
1&1&1&1&1&0&0&0&0&0&0&0&0&0&0&0&0&0&0&0\\
0&0&0&0&0&1&1&1&1&1&0&0&0&0&0&0&0&0&0&0\\
0&0&0&0&0&0&0&0&0&0&1&1&1&1&1&0&0&0&0&0\\
0&0&0&0&0&0&0&0&0&0&0&0&0&0&0&1&1&1&1&1\\
\hline
\al^4&\al^3&\al^2&\al&1&\al^4&\al^3&\al^2&\al&1&\al^4&\al^3&\al^2&\al&1&\al^4&\al^3&\al^2&\al&1\\
\al^8&\al^6&\al^4&\al^2&1&\al^8&\al^6&\al^4&\al^2&1&\al^8&\al^6&\al^4&\al^2&1&\al^8&\al^6&\al^4&\al^2&1\\
\al^4&\al^3&\al^2&\al&1&\al^3&\al^2&\al&1&\al^{-1}&\al^2&\al&1&\al^{-1}&\al^{-2}&\al&1&\al^{-1}&\al^{-2}&\al^{-3}\\
\al^8&\al^6&\al^4&\al^2&1&\al^7&\al^5&\al^3&\al&\al^{-1}&\al^6&\al^4&\al^2&1&\al^{-2}&\al^5&\al^3&\al&\al^{-1}&\al^{-3}\\
\end{array}
\right)
\end{eqnarray*}
assuming that $\al$ is an element in a finite field of order at least
5. For instance, we may take the finite field ${\rm GF}(8)$ and $\al$
a primitive root in ${\rm GF}(8)$, which has order 7.

Similarly, 
\begin{eqnarray*}
{\bf H} (5;(2,2,3,3))&=&
\left(
\begin{array}{rcl}
I_{4}&\otimes &H(2,5;0)\\
\hH(2,4;0)&\otimes & H(1,5;2)\\
\end{array}
\right),
\end{eqnarray*}
where now
\begin{eqnarray*}
H(2,5;0)\;\;\eq\;\; H_0&=&
\left(
\begin{array}{ccccc}
1&1&1&1&1\\
\al^4&\al^3&\al^2&\al&1\\
\end{array}
\right)
\end{eqnarray*}
and
\begin{eqnarray*}
H(1,5;2)\;\;\eq\;\; H_1&=&
\left(
\begin{array}{ccccc}
\al^8&\al^6&\al^4&\al^2&1\\
\end{array}
\right),
\end{eqnarray*}
giving explicitly ${\bf H} (5;(2,2,3,3))$ according to~(\ref{level2}) as
\begin{eqnarray*}
\footnotesize
\left(
\begin{array}{ccccc|ccccc|ccccc|ccccc}
1&1&1&1&1&0&0&0&0&0&0&0&0&0&0&0&0&0&0&0\\
\al^4&\al^3&\al^2&\al&1&0&0&0&0&0&0&0&0&0&0&0&0&0&0&0\\
0&0&0&0&0&1&1&1&1&1&0&0&0&0&0&0&0&0&0&0\\
0&0&0&0&0&\al^4&\al^3&\al^2&\al&1&0&0&0&0&0&0&0&0&0&0\\
0&0&0&0&0&0&0&0&0&0&1&1&1&1&1&0&0&0&0&0\\
0&0&0&0&0&0&0&0&0&0&\al^4&\al^3&\al^2&\al&1&0&0&0&0&0\\
0&0&0&0&0&0&0&0&0&0&0&0&0&0&0&1&1&1&1&1\\
0&0&0&0&0&0&0&0&0&0&0&0&0&0&0&\al^4&\al^3&\al^2&\al&1\\
\hline
\al^8&\al^6&\al^4&\al^2&1&\al^8&\al^6&\al^4&\al^2&1&\al^8&\al^6&\al^4&\al^2&1&\al^8&\al^6&\al^4&\al^2&1\\
\al^8&\al^6&\al^4&\al^2&1&\al^7&\al^5&\al^3&\al&
\al^{-1}&\al^6&\al^4&\al^2&1&\al^{-2}&\al^5&\al^3&\al&\al^{-1}&\al^{-3}\\
\end{array}
\right).
\end{eqnarray*}

As another example, take

\begin{eqnarray*}
{\bf H} (5;(2,2,4,4))&=&
\left(
\begin{array}{rcl}
I_{4}&\otimes & H(2,5;0)\\
\hH(2,4;0)&\otimes & H(2,5;2)\\
\end{array}
\right),
\end{eqnarray*}
where now
\begin{eqnarray*}
H(2,5;2)\;\;\eq\;\; H_1&=&
\left(
\begin{array}{ccccc}
\al^8&\al^6&\al^4&\al^2&1\\
\al^{12}&\al^9&\al^6&\al^3&1\\
\end{array}
\right),
\end{eqnarray*}
which gives, according to~(\ref{level2}), the following explicit value for ${\bf H} (5;(2,2,4,4))$:
\begin{eqnarray*}
\left(
{\footnotesize
\begin{array}{ccccc|ccccc|ccccc|ccccc}
1&1&1&1&1&0&0&0&0&0&0&0&0&0&0&0&0&0&0&0\\
\al^4&\al^3&\al^2&\al&1&0&0&0&0&0&0&0&0&0&0&0&0&0&0&0\\
0&0&0&0&0&1&1&1&1&1&0&0&0&0&0&0&0&0&0&0\\
0&0&0&0&0&\al^4&\al^3&\al^2&\al&1&0&0&0&0&0&0&0&0&0&0\\
0&0&0&0&0&0&0&0&0&0&1&1&1&1&1&0&0&0&0&0\\
0&0&0&0&0&0&0&0&0&0&\al^4&\al^3&\al^2&\al&1&0&0&0&0&0\\
0&0&0&0&0&0&0&0&0&0&0&0&0&0&0&1&1&1&1&1\\
0&0&0&0&0&0&0&0&0&0&0&0&0&0&0&\al^4&\al^3&\al^2&\al&1\\
\hline
\al^8&\al^6&\al^4&\al^2&1&\al^8&\al^6&\al^4&\al^2&1&\al^8&\al^6&\al^4&\al^2&1&\al^8&\al^6&\al^4&\al^2&1\\
\al^{12}&\al^9&\al^6&\al^3&1&\al^{12}&\al^9&\al^6&\al^3&1&\al^{12}&\al^9&\al^6&\al^3&1&\al^{12}&\al^9&\al^6&\al^3&1\\
\al^8&\al^6&\al^4&\al^2&1&\al^7&\al^5&\al^3&\al&\al^{-1}&\al^6&\al^4&\al^2&1&\al^{-2}&\al^{5}&\al^3&\al&\al^{-1}&\al^{-3}\\
\al^{12}&\al^9&\al^6&\al^3&1&\al^{11}&\al^8&\al^5&\al^2&\al^{-1}&\al^{10}&\al^7&\al^4&\al&\al^{-2}&
\al^9&\al^6&\al^3&1&\al^{-3}\\
\end{array}
}
\right).
\end{eqnarray*}

\qed
}
\end{ex}

\begin{ex}
\label{ex3}
{\em
Assume now $t\eq 3$, i.e., $\uu\eq
\left(\overbrace{u_0,u_0,\ldots,u_0}^{s_0},\overbrace{u_1,u_1,\ldots,u_1}^{s_1},
\overbrace{u_2,u_2,\ldots,u_2}^{s_2}\right)$
and
$\C(n;\uu)$ is a 3-level GC code..
Then, according to~(\ref{eqmain}) and~(\ref{eqmain2}),

\begin{eqnarray}
\nonumber
{\bf H} (n;\uu) &=&
\left(
\begin{array}{rcl}
I_{s_0+s_1+s_2}&\otimes & H(u_0,n;0)\\
\hH(s_{2},m;0)&\otimes & H(u_{2}-u_0,n;u_0)\\
\hH(s_{1},m;s_{2})&\otimes & H(u_{1}-u_0,n;u_0)\\
\end{array}
\right)\\
\label{level3}
&=&
\left(
\begin{array}{c|c|c|c}
H_0&\uzero &\ldots &\uzero\\
\uzero&H_0&\ldots &\uzero\\
\vdots &\vdots &\ddots &\vdots\\
\uzero&\uzero&\ldots &H_0\\
\hline
H_{2}&H_{2}&\ldots
&H_{2}\\
H_{2}&\al^{-1}H_{2}&\ldots
&\al^{-(m-1)}H_{2}\\
\vdots &\vdots &\ddots &\vdots\\
H_{2}&\al^{-(s_{1}-1)}H_{2}&\ldots
&\al^{-(m-1)(s_{1}-1)}H_{2}\\
\hline
H_{1}&\al^{-s_{1}}H_{1}&\ldots &\al^{-(m-1)s_{1}}H_{1}\\
H_{1}&\al^{-(s_{1}+1)}H_{1}&\ldots
&\al^{-(m-1)(s_{1}+1)}H_{1}\\
\vdots &\vdots &\ddots &\vdots\\
H_{1}&\al^{-(s_{1}+s_2-1)}H_{1}&\ldots
&\al^{-(m-1)(s_{1}+s_2-1)}H_{1}\\
\end{array}
\right).
\end{eqnarray}

If we take
$\uu\eq (1,1,2,3)$, then the parity-check matrix
of the 3-level code
$\C(5;(1,1,2,3))$, is given by

\begin{eqnarray*}
 {\bf H} (5;(1,1,2,3))&=&
\left(
\begin{array}{rcl}
I_{4}&\otimes & H(1,5;0)\\
\hH(1,4;0)&\otimes & H(2,5;1)\\
\hH(1,4;1)&\otimes & H(1,5;1)\\
\end{array}
\right),
\end{eqnarray*}
which explicitly gives, according to~(\ref{level3}),
\begin{eqnarray*}
\left(
\begin{array}{ccccc|ccccc|ccccc|ccccc}
1&1&1&1&1&0&0&0&0&0&0&0&0&0&0&0&0&0&0&0\\
0&0&0&0&0&1&1&1&1&1&0&0&0&0&0&0&0&0&0&0\\
0&0&0&0&0&0&0&0&0&0&1&1&1&1&1&0&0&0&0&0\\
0&0&0&0&0&0&0&0&0&0&0&0&0&0&0&1&1&1&1&1\\
\hline
\al^4&\al^3&\al^2&\al&1&\al^4&\al^3&\al^2&\al&1&\al^4&\al^3&\al^2&\al&1&\al^4&\al^3&\al^2&\al&1\\
\al^{8}&\al^6&\al^4&\al^2&1&\al^{8}&\al^6&\al^4&\al^2&1&\al^8&\al^6&\al^4&\al^2&1&\al^8&\al^6&\al^4&\al^2&1\\
\hline
\al^4&\al^3&\al^2&\al&1&\al^3&\al^2&\al&1&\al^{-1}&\al^2&\al&1&\al^{-1}&\al^{-2}&\al&1&\al^{-1}&\al^{-2}&\al^{-3}\\
\end{array}
\right),
\end{eqnarray*}
while if we take
$\uu\eq (1,2,2,3)$, then the parity-check matrix
${\bf H} (5;(1,2,2,3))$ of the 3-level code
$\C(5;(1,2,2,3))$, is given by
\begin{eqnarray*}
 {\bf H} (5;(1,2,2,3))&=&
\left(
\begin{array}{rcl}
I_{4}&\otimes &H(1,5;0)\\
\hH(1,4;0)&\otimes &H(2,5;1)\\
\hH(2,4;1)&\otimes &H(1,5;1)\\
\end{array}
\right),
\end{eqnarray*}
which explicitly gives, according to~(\ref{level3}),
{\footnotesize
\begin{eqnarray*}
\left(
\begin{array}{ccccc|ccccc|ccccc|ccccc}
1&1&1&1&1&0&0&0&0&0&0&0&0&0&0&0&0&0&0&0\\
0&0&0&0&0&1&1&1&1&1&0&0&0&0&0&0&0&0&0&0\\
0&0&0&0&0&0&0&0&0&0&1&1&1&1&1&0&0&0&0&0\\
0&0&0&0&0&0&0&0&0&0&0&0&0&0&0&1&1&1&1&1\\
\hline
\al^4&\al^3&\al^2&\al&1&\al^4&\al^3&\al^2&\al&1&\al^4&\al^3&\al^2&\al&1&\al^4&\al^3&\al^2&\al&1\\
\al^{8}&\al^6&\al^4&\al^2&1&\al^{8}&\al^6&\al^4&\al^2&1&\al^8&\al^6&\al^4&\al^2&1&\al^8&\al^6&\al^4&\al^2&1\\
\hline
\al^4&\al^3&\al^2&\al&1&\al^3&\al^2&\al&1&\al^{-1}&\al^2&\al&1&\al^{-1}&\al^{-2}&\al&1&\al^{-1}&\al^{-2}&\al^{-3}\\
\al^4&\al^3&\al^2&\al&1&\al^2&\al&1&\al^{-1}&\al^{-2}
&1&\al^{-1}&\al^{-2}&\al^{-3}&\al^{-4}&\al^{-2}&\al^{-3}&\al^{-4}&\al^{-5}&\al^{-6}\\
\end{array}
\right),
\end{eqnarray*}
}
again assuming that $\al$ is an element in a finite field with
order at least 5.

\qed
}
\end{ex}

We give next the main property of $t$-level GC codes.

\begin{theo}
\label{theo1}
{\em
Consider the integers $n\leq 2^b-1$, $t\geq 1$, $s_i\geq 1$ for
$0\leq i\leq t-1$ and $1\leq u_0<u_1<\ldots <u_{t-1}\leq n-1$.
Let $m\eq s_0+s_1+\cdots +s_{t-1}$ and $\uu$ be given
by~(\ref{equu}). Then
the $t$-level GC code $\C(n;\uu)$  whose parity-check matrix ${\bf
H}(n;\uu)$ is given by~(\ref{eqmain}) can correct up to
$u_i$ erasures in any $s_i$ rows, $0\leq i\leq t-1$, of an $m\times n$ array
corresponding to a codeword in $\C(n;\uu)$.
}
\end{theo}

Theorem~\ref{theo1} will be proved in Section~\ref{encdec}, where we
will show that there is a decoding algorithm correcting the erasure
instances described in the theorem. Next we illustrate it with an
example.

\begin{ex}
\label{extheo1}
{\rm
Consider code $\C(5;(1,1,3,3))$ given in Example~\ref{ex2} corresponding to
$4\times 5$ arrays. According to Theorem~\ref{theo1},
up to three erasures will be corrected in any pair of rows,
while the remaining rows can correct up to one erasure.
For example, denoting erasures by $X$, the following arrays are
correctable in $\C(5;(1,1,3,3))$:

$$
\begin{array}{cc}
\begin{array}{|c|c|c|c|c|}
\hline
\phantom{X}&X&\phantom{X}&\phantom{X}&\phantom{X}\\
\hline
X&\phantom{X}&\phantom{X}&X&X\\
\hline
\phantom{X}&\phantom{X}&\phantom{X}&\phantom{X}&X\\
\hline
\phantom{X}&X&X&X&\phantom{X}\\
\hline
\end{array}&
\begin{array}{|c|c|c|c|c|}
\hline
X&X&\phantom{X}&\phantom{X}&X\\
\hline
X&\phantom{X}&\phantom{X}&\phantom{X}&\phantom{X}\\
\hline
\phantom{X}&X&\phantom{X}&\phantom{X}&\phantom{X}\\
\hline
\phantom{X}&X&\phantom{X}&X&X\\
\hline
\end{array}
\end{array}
$$

A way to correct the erasures above is by using the parity-check
matrix ${\bf H} (5;(1,1,3,3))$ of the code given in Example~\ref{ex2}:
syndromes are computed, and first the rows that experienced one
erasure are corrected (using single parity). Once
they are corrected, the syndromes are updated.
To correct the two rows with 3 erasures each, it is needed to solve a
linear system of 6 equations with 6 unknowns, which can be easily
done, for instance, by Gaussian elimination (we will present a much
more efficient decoding algorithm in Section~\ref{encdec}).

As is the case in general with erasure decoding, encoding is a
special case of decoding. For example, for $\C(5;(1,1,3,3))$, we may
choose to place the parities at the end of each row, like below, in
either increasing or decreasing order on the number of erasures (the
STAIR codes~\cite{ll} use such an encoding ordering):

$$
\begin{array}{cc}
\begin{array}{|c|c|c|c|c|}
\hline
\phantom{X}&\phantom{X}&\phantom{X}&\phantom{X}&X\\
\hline
\phantom{X}&\phantom{X}&\phantom{X}&\phantom{X}&X\\
\hline
\phantom{X}&\phantom{X}&X&X&X\\
\hline
\phantom{X}&\phantom{X}&X&X&X\\
\hline
\end{array}&
\begin{array}{|c|c|c|c|c|}
\hline
\phantom{X}&\phantom{X}&X&X&X\\
\hline
\phantom{X}&\phantom{X}&X&X&X\\
\hline
\phantom{X}&\phantom{X}&\phantom{X}&\phantom{X}&X\\
\hline
\phantom{X}&\phantom{X}&\phantom{X}&\phantom{X}&X\\
\hline
\end{array}
\end{array}
$$

Knowing a priori the erased entries allows for shortcuts in the
processing time by precomputing certain operations. We will give some
details in Section~\ref{encdec}.

Similarly, $\C(5;(1,1,2,3))$ corresponds to a $4\times 5$ array
such that one row can correct up to three erasures, one of the
remaining three rows can correct up to two erasures, and the
remaining rows can correct up to one erasure.
For example, the following arrays are
correctable in $\C(5;(1,1,2,3))$:

$$
\begin{array}{cc}
\begin{array}{|c|c|c|c|c|}
\hline
\phantom{X}&X&\phantom{X}&\phantom{X}&\phantom{X}\\
\hline
X&\phantom{X}&\phantom{X}&X&X\\
\hline
\phantom{X}&\phantom{X}&\phantom{X}&\phantom{X}&X\\
\hline
\phantom{X}&X&&X&\phantom{X}\\
\hline
\end{array}&
\begin{array}{|c|c|c|c|c|}
\hline
&X&\phantom{X}&\phantom{X}&X\\
\hline
X&\phantom{X}&\phantom{X}&\phantom{X}&\phantom{X}\\
\hline
\phantom{X}&X&\phantom{X}&\phantom{X}&\phantom{X}\\
\hline
\phantom{X}&X&\phantom{X}&X&X\\
\hline
\end{array}
\end{array}
$$

Let us examine more closely the array on the left above. Consider its
parity-check matrix ${\bf H} (5;(1,1,2,3))$ as given in
Example~\ref{ex3}. The rows with only one erasure are corrected using
single parity, so we are left with the array

$$
\begin{array}{|c|c|c|c|c|}
\hline
\phantom{X}&\phantom{X}&\phantom{X}&\phantom{X}&\phantom{X}\\
\hline
X&\phantom{X}&\phantom{X}&X&X\\
\hline
\phantom{X}&\phantom{X}&\phantom{X}&\phantom{X}&\phantom{X}\\
\hline
\phantom{X}&X&&X&\phantom{X}\\
\hline
\end{array}
$$

By writing the array as a vector row-wise, the erased entries
correspond to locations 5, 8, 9, 16 and 18. The $5\times 5$ matrix
from ${\bf H} (5;(1,1,2,3))$ corresponding to these locations is

\begin{eqnarray*}
 \tH&=&
\left(
\begin{array}{ccc|cc}
1&1&1&0&0\\
0&0&0&1&1\\
\hline
\al^4&\al&1&\al^3&\al\\
\al^{8}&\al^2&1&\al^6&\al^2\\
\hline
\al^3&1&\al^{-1}&1&\al^{-2}\\
\end{array}
\right),
\end{eqnarray*}
which we must prove is invertible. 

To see this, let $H_0(3)\eq
(1\;1\;1)$, $H_0(2)\eq (1\;1)$, 
$H_2(3)\eq \left(
\begin{array}{ccc}
\al^4&\al&1\\
\al^8&\al^2&1\\
\end{array}
\right)$,
$H_2(2)\eq \left(
\begin{array}{cc}
\al^3&\al\\
\al^6&\al^2\\
\end{array}
\right)$,
$H_1(3)\eq (\al^4\;\al\;1)$ and
$H_1(2)\eq (\al^3\;\al)$. Then, we can write $\tH$ as

\begin{eqnarray*}
 \tH&=&
\left(
\begin{array}{c|c}
H_0(3)&\uzero\\
\uzero &H_0(2)\\
\hline
H_2(3)&H_2(2)\\
\hline
\al^{-1}H_1(3)&\al^{-3}H_1(2)\\
\end{array}
\right).
\end{eqnarray*}

Since $$\left(\begin{array}{cc}
1&1\\
\al^{-1}&\al^{-3}\\
\end{array}\right)$$ is a Vandermonde matrix, in particular it is
invertible and there is a linear combination of its rows that
transforms it into an upper triangular matrix with 1s in the
diagonal, i.e., $\left(\begin{array}{cc}
1&1\\
0&1\\
\end{array}\right)$. Notice that since $H_1(3)$ (resp. $H_1(2)$) corresponds to the
first row of $H_2(3)$ (resp. $H_2(2)$), we can apply this linear combination to the rows of $\tH$
corresponding to $H_1(3)$ and $H_1(2)$, so we obtain

\begin{eqnarray*}
 \tH'&=&
\left(
\begin{array}{c|c}
H_0(3)&\uzero\\
\uzero &H_0(2)\\
\hline
H_2(3)&H_2(2)\\
\hline
\uzero&H_1(2)\\
\end{array}
\right).
\end{eqnarray*}
Permuting the rows of $\tH'$, we have

\begin{eqnarray*}
 \tH''&=&
\left(
\begin{array}{c|c}
H_0(3)&\uzero\\
H_2(3)&H_2(2)\\
\hline
\uzero &H_0(2)\\
\uzero&H_1(2)\\
\end{array}
\right).
\end{eqnarray*}

By properties of determinants, the determinant of $\tH''$ is the
product of the determinants of
\begin{eqnarray*}
\left(
\begin{array}{c}
H_0(3)\\
H_2(3)\\
\end{array}
\right)&=&
\left(
\begin{array}{ccc}
1&1&1\\
\al^4&\al&1\\
\al^8&\al^2&1\\
\end{array}
\right)
\end{eqnarray*}
and
\begin{eqnarray*}
\left(
\begin{array}{c}
H_0(2)\\
H_1(2)\\
\end{array}
\right)&=&
\left(
\begin{array}{cc}
1&1\\
\al^3&\al\\
\end{array}
\right).
\end{eqnarray*}
Since these determinants are both Vandermonde determinants they are
non-zero, thus, their product is non-zero.

\qed
}
\end{ex}

The decoding algorithm proving Theorem~\ref{theo1} to be presented in
the next section develops the idea presented in Example~\ref{extheo1}.

The following result was given without proof in~\cite{tk}:

\begin{cor}
\label{cor1}
{\em
Consider the $t$-level GC code $\C(n;\uu)$ of
Theorem~\ref{theo1}. Then, if $\hs_t\eq 0$ and $\hs_i\eq
\sum_{j=i}^t s_j$ for $0\leq i\leq t-1$, the minimum
distance of $\C(n;\uu)$ is given by

\begin{eqnarray*}
d&=&\min\left\{\left(\hs_{i+1}+1\right)\left(u_i+1\right)\;,\;0\leq
i\leq t-1\right\}.
\end{eqnarray*}
}
\end{cor}

\noindent\pf 
Assume that there is a codeword that has exactly $\hs_{i+1}$ rows of
weight $u_i+1$ and one row of weight $u_i$, while all the other rows
are zero (notice that when $i\eq t-1$, this simply means that there
is a codeword
consisting of a row of weight $u_{t-1}$, while all the other rows are
zero). By Theorem~\ref{theo1}, such a codeword would be corrected by
the code as the zero codeword, thus

\begin{eqnarray*}
d&> &\min\left\{\left(\hs_{i+1}\right)\left(u_i+1\right)+u_i\;,\;0\leq
i\leq t-1\right\},
\end{eqnarray*}
or,
\begin{eqnarray*}
d&\geq &\min\left\{\left(\hs_{i+1}+1\right)\left(u_i+1\right)\;,\;0\leq
i\leq t-1\right\}.
\end{eqnarray*}

In order to show equality, we need to prove that for each $i$, $1\leq
i\leq t-1$, there is a codeword in $\C(n;\uu)$ of weight
$\left(\hs_{i+1}+1\right)\left(u_i+1\right)$.

Consider first the case $i\eq t-1$, thus, we have to prove that there
is a codeword of weight $u_{t-1}+1$. Let $\uu$ be a codeword of
weight $u_{t-1}+1$ in the $[n,n-u_{t-1},u_{t-1}+1]$ RS code whose
parity-check matrix is given by $H(u_{t-1},n;0)$, and $\uzero_n$ a
zero vector of length $n$. Then, according to~(\ref{eqmain}) and~(\ref{eqmain2}), vector
$$(\uu,\overbrace{\uzero_n,\uzero_n,\ldots,\uzero_n}^{m-1})$$ is a codeword in
$\C(n;\uu)$ of weight $u_{t-1}+1$ (notice that the rows of $H_i$ in~(\ref{eqmain2})
are contained in the rows of $H(u_{t-1},n;0)$ for
$0\leq i\leq t-1$).

Next consider $0\leq i\leq t-2$. Let $\uu$ be a codeword of
weight $u_i+1$ in the $[n,n-u_{i},u_{i}+1]$ RS code whose
parity-check matrix is given by $H(u_{i},n;0)$. 
Notice that the rows of $H_j$ in~(\ref{eqmain2}) are contained in the
rows of $H(u_{i},n;0)$ for $0\leq j\leq i$.
Let $\uv$ be a
codeword of weight $\hs_{i+1}+1$ in the RS code whose
parity-check matrix is given by
$\hat{H}(\hs_{i+1},\hs_{i+1}+1;0)$. 
Explicitly, let $\uv\eq (v_0,v_1,\ldots,v_{\hs_{i+1}})$.

Consider the following vector of length $mn$:
\begin{eqnarray*}
\uw&=&\left(v_0\,\uu\;,\;v_1\,\uu\;,\;\ldots\;,\;v_{\hs_{i+1}}\,\uu\;,\;\uzero\right),
\end{eqnarray*}
where $\uzero$ is a vector of length $n\left(\sum_{j=0}^is_j\right)$.
According to~(\ref{eqmain}) and~(\ref{eqmain2}), we have to
show that vector $\uw$ is a codeword in $\C(n;\uu)$. Certainly, since
$H(u_{i},n;0)\uu^T\eq \uzero$, we have that, according
to~(\ref{eqmain2}), the inner product of the 
rows of ${\bf H} (n;\uu)$ involving $H_j$, $0\leq j\leq i$, with $\uw$ are
zero. 

On the other hand, take any of the rows of ${\bf H} (n;\uu)$
involving $H_j$, $i+1\leq j\leq t-1$, 
in~(\ref{eqmain2}). The inner
product of such a row with $\uw$ is also zero, since it is a constant
times the inner product of $\uv$ with a row of the parity-check matrix
$\hat{H}(\hs_{i+1},\hs_{i+1}+1;0)$, which is zero by construction.
\qed

The following example illustrates Corollary~\ref{cor1} and its proof.

\begin{ex}
\label{excor1}
{\rm
Consider
code $\C(5;(1,2,2,3))$ as given in
Example~\ref{ex3}. Corollary~\ref{cor1} states that the minimum
distance of $\C(5;(1,2,2,3))$ is given by

\begin{eqnarray*}
d&=&\min\left\{(4)(2)\;,\;(2)(3)\;,\;4
\right\}\;\;=\;\;4.
\end{eqnarray*}

Certainly there are no codewords of weight 3. According to~(\ref{eqmain2}), the parity-check matrix
${\bf H}(5;(1,2,2,3))$ is given by

\begin{eqnarray*}
 {\bf H} (5;(1,2,2,3))&=&
\left(
\begin{array}{c|c|c|c}
H_0&\uzero&\uzero&\uzero\\
\uzero&H_0&\uzero&\uzero\\
\uzero&\uzero&H_0&\uzero\\
\uzero&\uzero&\uzero&H_0\\
\hline
H_2&H_2&H_2&H_2\\
\hline
H_1&\al^{-1}H_1&\al^{-2}H_1&\al^{-3}H_1\\
H_1&\al^{-2}H_1&\al^{-4}H_1&\al^{-6}H_1\\
\end{array}
\right),
\end{eqnarray*}
where
$$H_0\eq\left(\begin{array}{ccccc}1&1&1&1&1\\\end{array}\right),$$
$$H_1\eq\left(\begin{array}{ccccc}\al^4&\al^3&\al^2&\al&1\\\end{array}\right)$$
and
$$H_2\eq\left(\begin{array}{ccccc}\al^4&\al^3&\al^2&\al&1\\\al^8&\al^6&\al^4&\al^2&1\\\end{array}\right).$$

Consider the [5,2,4] RS code whose parity-check matrix is
$\left({H_0\atop H_2}\right)$. Let $\uu$ be a codeword of weight 4 in
such a code. Then, $(\uu\,,\,\uzero\,,\,\uzero\,,\,\uzero)$ is a
codeword of weight 4 in
$\C(5;(1,2,2,3))$, since we easily see that its inner product with
the rows of ${\bf H} (5;(1,2,2,3))$ is zero.

Let us show next the existence of a codeword of weight $(2)(3)\eq 6$ with
two non-zero rows of weight 3. Take a codeword $\uu$ of weight 3 in the
$[5,3,3]$ code whose parity-check matrix is given by $\left({H_0\atop
H_1}\right)$.
Consider a codeword of weight 2 in the $[5,4,2]$ code whose parity-check matrix
is $\left(\begin{array}{cccc}1&1&1&1\\\end{array}\right)$, say,
(1,1,0,0). Then, we can see that $\uw\eq (\uu,\uu,\uzero,\uzero)$ is a
codeword of weight (2)(3) in $\C(5;(1,2,2,3))$. In effect, the inner
product of $\uw$ with the first 5 and the last 2 rows of ${\bf H}
(5;(1,2,2,3))$ is zero, since the inner product of the rows of $H_0$
and of $H_1$ with $\uu$ are zero by construction. Now, if the inner
product of $\uu$ with the second row of $H_2$ is, say, $\ga$, then the inner
product of $\uw$ with the sixth row of ${\bf H} (5;(1,2,2,3))$ is $\ga\xor\ga\eq 0$.

Finally, let us show that there is a codeword of weight $(4)(2)\eq 8$, with
four non-zero rows of weight 2. Take a codeword $\uu$ of weight 2 in the
$[5,4,2]$ code whose parity-check matrix is given by $H_0$, for
instance, $\uu\eq (1,1,0,0,0)$ is such a codeword. Take a codeword $\uv\eq
(v_0,v_1,v_2,v_3)$ of weight 4 in the $[4,1,4]$ code whose
parity-check matrix is
\begin{eqnarray*}
\hH(3,4;0) &=&
\left(
\begin{array}{cccc}
1&1&1&1\\
1&\al^{-1}&\al^{-2}&\al^{-3}\\
1&\al^{-2}&\al^{-4}&\al^{-6}\\
\end{array}
\right).
\end{eqnarray*}
Take $\uw\eq \left(v_0\,\uu\;,\;v_1\,\uu\;,\;v_2\,\uu\;,\;v_3\uu\right)$. Then, $\uw$
is a codeword of weight (4)(2) in $\C(5;(1,2,2,3))$. In effect, the
inner product of $\uw$ with any of the first four rows of ${\bf H}
(5;(1,2,2,3))$ is zero, since the inner product of $\uu$ with the row
of $H_0$ is zero. Next take any of the remaining rows, and assume
that the inner product of $\uu$ with the first 5 coordinates of
such row is $\ga$. Then the inner product of $\uw$ with the row is
given by $\ga$ times the inner product of $\uv$ with a row of
$\hH(3,4;0)$, which is zero by construction.

\qed
}
\end{ex}

\section{Encoding and Decoding}
\label{encdec}
In erasure decoding, encoding is a special case of the decoding. The
decoding algorithm to be presented next also proves
Theorem~\ref{theo1}. 

Assume that we have a $t$-level GC-code $\C(n;\uu)$ as given by
Definition~\ref{defGC}. The codewords are $m\times n$ arrays. As
before, let $\uu$ be given by~(\ref{equu}), ${\bf v}$ be a received
$m\times n$ array with erasures, and without loss of
generality, assume that there are $s_{t-1}$ rows of ${\bf v}$ with
$u_{t-1}$ erasures each, $s_{t-2}$ rows of ${\bf v}$ with
$u_{t-2}$ erasures each, and so on, until finally there are $s_0$
rows of ${\bf v}$ with $u_0$ erasures each.
Let $\si :\{0,1,\ldots,m-1\}\;\ra\;\{0,1,\ldots,m-1\}$ be a permutation
of the rows of ${\bf v}$ and ${\bf v}_{\si}$ the array ${\bf v}$ with the rows
permuted according to $\si$, such that the first $s_{t-1}$ rows of ${\bf v}_{\si}$ have
$u_{t-1}$ erasures each, the next $s_{t-2}$ rows of ${\bf v}_{\si}$ have
$u_{t-2}$ erasures each, and so on, until finally the last $s_0$
rows of ${\bf v}_{\si}$ have $s_0$ erasures each.

We permute accordingly the columns of the parity-check
matrix ${\bf H} (n;\uu)$ of $\C(n;\uu)$ to give the permuted
parity-check matrix ${\bf H}_{\si}(n;\uu)$ corresponding to a
permuted code $\C_{\sigma}(n;\uu)$. Specifically, if we write the
$\left(\sum_{i=0}^{t-1}u_is_i\right)\times mn$ parity-check matrix
${\bf H} (n;\uu)$ as

\begin{eqnarray}
\label{PCH}
{\bf H} (n;\uu)&=&
\left(\begin{array}{cccc}
{\bf H}_0&{\bf H}_1&\ldots &{\bf H}_{m-1}\\
\end{array}\right),
\end{eqnarray}
where each $\bf H_i$ is a $\left(\sum_{i=0}^{t-1}u_is_i\right)\times
n$ matrix, and let $i_0,i_1,\ldots, i_{m-1}$ be such that
$\si(i_j)\eq j$ for $0\leq j\leq m-1$, then

\begin{eqnarray}
\label{PCHp}
{\bf H}_{\si}(n;\uu)&=&
\left(\begin{array}{cccc}
{\bf H}_{i_0}&{\bf H}_{i_1}&\ldots &{\bf H}_{i_{m-1}}\\ 
\end{array}\right)
\end{eqnarray}
and $\C_{\sigma}(n;\uu)$ is the permuted code given by the parity-check matrix
${\bf H}_{\si}(n;\uu)$. We will describe next how to use this permuted
parity-check matrix in order to implement an efficient decoding algorithm.

Based on ${\bf H}(n;\uu)$ as given by~(\ref{eqmain2}), ${\bf
H}_{\si}(n;\uu)$ is given by 

\begin{eqnarray}
\label{eqmain3}
{\bf H}_{\sigma}(n;\uu) &=&
\left(
\begin{array}{c|c|c|c}
H_0&\uzero &\ldots &\uzero\\
\uzero&H_0&\ldots &\uzero\\
\vdots &\vdots &\ddots &\vdots\\
\uzero&\uzero&\ldots &H_0\\
\hline
H_{t-1}&H_{t-1}&\ldots
&H_{t-1}\\
\al^{-i_0}H_{t-1}&\al^{-i_1}H_{t-1}&\ldots
&\al^{-i_{m-1}}H_{t-1}\\
\vdots &\vdots &\ddots &\vdots\\
\al^{-i_0(\hs_{t-1}-1)}H_{t-1}&\al^{-i_1(\hs_{t-1}-1)}H_{t-1}&\ldots
&\al^{-i_{m-1}(\hs_{t-1}-1)}H_{t-1}\\
\hline
\al^{-i_0\hs_{t-1}}H_{t-2}&\al^{-i_1\hs_{t-1}}H_{t-2}&\ldots
&\al^{-i_{m-1}\hs_{t-1}}H_{t-2}\\
\al^{-i_0(\hs_{t-1}+1)}H_{t-2}&\al^{-i_1(\hs_{t-1}+1)}H_{t-2}&\ldots
&\al^{-i_{m-1}(\hs_{t-1}+1)}H_{t-2}\\
\vdots &\vdots &\ddots &\vdots\\
\al^{-i_0(\hs_{t-2}-1)}H_{t-2}&\al^{-i_1(\hs_{t-2}-1)}H_{t-2}&\ldots
&\al^{-i_{m-1}(\hs_{t-2}-1)}H_{t-2}\\
\hline
\vdots &\vdots &\ddots &\vdots\\
\hline
\al^{-i_0\hs_{i+1}}H_{i}&\al^{-i_1\hs_{i+1}}H_{i}&\ldots
&\al^{-i_{m-1}\hs_{i+1}}H_{i}\\
\al^{-i_0(\hs_{i+1}+1)}H_{i}&\al^{-i_1(\hs_{i+1}+1)}H_{i}&\ldots
&\al^{-i_{m-1}(\hs_{i+1}+1)}H_{i}\\
\vdots &\vdots &\ddots &\vdots\\
\al^{-i_0(\hs_i-1)}H_{i}&\al^{-i_1(\hs_i-1)}H_{i}&\ldots
&\al^{-i_{m-1}(\hs_i-1)}H_{i}\\
\hline
\vdots &\vdots &\ddots &\vdots\\
\hline
\al^{-i_0\hs_2}H_{1}&\al^{-i_1\hs_2}H_{1}&\ldots
&\al^{-i_{m-1}\hs_2}H_{1}\\
\al^{-i_0(\hs_2+1)}H_{1}&\al^{-i_1(\hs_2+1)}H_{1}&\ldots
&\al^{-i_{m-1}(\hs_2+1)}H_{1}\\
\vdots &\vdots &\ddots &\vdots\\
\al^{-i_0(\hs_1-1)}H_{1}&\al^{-i_1(\hs_1-1)}H_{1}&\ldots
&\al^{-i_{m-1}(\hs_1-1)}H_{1}\\
\end{array}
\right)
\end{eqnarray}

Consider next the $\hs_1\times m$ matrix

\begin{eqnarray}
\label{eqmain5}
{\bf \hat{H}}_{\sigma}(m;\uu) &=&
\left(
\begin{array}{cccc}
1&1&\ldots &1\\
\al^{-i_0}&\al^{-i_1}&\ldots
&\al^{-i_{m-1}}\\
\vdots &\vdots &\ddots &\vdots\\
\al^{-i_0(\hs_{t-1}-1)}&\al^{-i_1(\hs_{t-1}-1)}&\ldots
&\al^{-i_{m-1}(\hs_{t-1}-1)}\\
\hline
\al^{-i_0\hs_{t-1}}&\al^{-i_1\hs_{t-1}}&\ldots
&\al^{-i_{m-1}\hs_{t-1}}\\
\al^{-i_0(\hs_{t-1}+1)}&\al^{-i_1(\hs_{t-1}+1)}&\ldots
&\al^{-i_{m-1}(\hs_{t-1}+1)}\\
\vdots &\vdots &\ddots &\vdots\\
\al^{-i_0(\hs_{t-2}-1)}&\al^{-i_1(\hs_{t-2}-1)}&\ldots
&\al^{-i_{m-1}(\hs_{t-2}-1)}\\
\hline
\vdots &\vdots &\ddots &\vdots\\
\hline
\al^{-i_0\hs_{i+1}}&\al^{-i_1\hs_{i+1}}&\ldots
&\al^{-i_{m-1}\hs_{i+1}}\\
\al^{-i_0(\hs_{i+1}+1)}&\al^{-i_1(\hs_{i+1}+1)}&\ldots
&\al^{-i_{m-1}(\hs_{i+1}+1)}\\
\vdots &\vdots &\ddots &\vdots\\
\al^{-i_0(\hs_i-1)}&\al^{-i_1(\hs_i-1)}&\ldots
&\al^{-i_{m-1}(\hs_i-1)}\\
\hline
\vdots &\vdots &\ddots &\vdots\\
\hline
\al^{-i_0\hs_2}&\al^{-i_1\hs_2}&\ldots
&\al^{-i_{m-1}\hs_2}\\
\al^{-i_0(\hs_2+1)}&\al^{-i_1(\hs_2+1)}&\ldots
&\al^{-i_{m-1}(\hs_2+1)}\\
\vdots &\vdots &\ddots &\vdots\\
\al^{-i_0(\hs_1-1)}&\al^{-i_1(\hs_1-1)}&\ldots
&\al^{-i_{m-1}(\hs_1-1)}\\
\end{array}
\right)
\end{eqnarray}

Since ${\bf \hat{H}}_{\sigma}(m;\uu)$ as given by~(\ref{eqmain5}) is
a (rectangular) Vandermonde matrix and 
$\hs_1<\hs_0\eq m$, there is a linear combination that transforms the
matrix ${\bf \hat{H}}_{\sigma}(m;\uu)$ above into an upper triangular form (for instance, by doing
Gaussian elimination). Specifically, let the upper triangular form be

\begin{eqnarray}
\label{triang}
\left(
\begin{array}{ccccccccc}
1&1&1&\ldots &1&1&\ldots &1&1\\
0&1&\ga_{1,2}&\ldots &\ga_{1,\us_1-2}&\ga_{1,\us_1-1}&\ldots &\ga_{1,m-2}&\ga_{1,m-1}\\
0&0&1&\ldots &\ga_{2,\us_1-2}&\ga_{2,\us_1-1}&\ldots &\ga_{2,m-2}&\ga_{2,m-1}\\
\vdots&\vdots&\vdots&\ddots&\vdots&\vdots&\ddots&\vdots&\vdots\\
0&0&0&\ldots &1&\ga_{\us_1-2,\us_1-1}&\ldots
&\ga_{\us_1-2,m-2}&\ga_{\us_1-2,m-1}\\
0&0&0&\ldots &0&1&\ldots &\ga_{\us_1-1,m-2}&\ga_{\us_1-1,m-1}\\
\end{array}
\right)
\end{eqnarray}

Since the rows of $H_j$ are contained in the rows of $H_i$ when
$1\leq j<i$, by applying the linear combination that transforms ${\bf
\hat{H}}_{\sigma}(m;\uu)$ into this upper triangular matrix 
given by~(\ref{triang}) to the last $\sum_{j=1}^{t-1}s_j(u_j-u_0)$ rows of ${\bf H}_{\sigma}(n;\uu)$ as
given by~(\ref{eqmain3}),
we obtain

\begin{eqnarray}
\label{triangb}
\left(
\begin{array}{ccccccccc}
H_{t-1}&H_{t-1}&H_{t-1}&\ldots &H_{t-1}&H_{t-1}&\ldots &H_{t-1}&H_{t-1}\\
0&H_{t-1}&\ga_{1,2}H_{t-1}&\ldots &\ga_{1,\us_1-2}H_{t-1}&\ga_{1,\us_1-1}H_{t-1}&\ldots &\ga_{1,m-2}H_{t-1}&\ga_{1,m-1}H_{t-1}\\
0&0&H_{t-1}&\ldots &\ga_{2,\us_1-2}H_{t-1}&\ga_{2,\us_1-1}H_{t-1}&\ldots &\ga_{2,m-2}H_{t-1}&\ga_{2,m-1}H_{t-1}\\
\vdots&\vdots&\vdots&\ddots&\vdots&\vdots&\ddots&\vdots&\vdots\\
0&0&0&\ldots &H_1&\ga_{\us_1-2,\us_1-1}H_1&\ldots &\ga_{\us_1-2,m-2}H_1&\ga_{\us_1-2,m-1}H_1\\
0&0&0&\ldots &0&H_1&\ldots &\ga_{\us_1-1,m-2}H_1&\ga_{\us_1-1,m-1}H_1\\
\end{array}
\right)
\end{eqnarray}


Combining the first $s_0u_0$ rows of ${\bf H}_{\sigma}(n;\uu)$ as
given by~(\ref{eqmain3}) with the matrix given by~(\ref{triangb}),
after some rearrangement of the rows, we obtain the pseudo upper-triangular matrix
$\overset{\tiny\tr}{\bf H}_{\sigma}(n;\uu)$ given by~(\ref{eqmain4})
below: 

\begin{eqnarray}
\label{eqmain4}
\left(
\begin{array}{c|c|c|c|c|c|c|c}
H_0&\uzero &\ldots &\uzero&\uzero&\ldots &\uzero&\uzero\\
H_{t-1}&H_{t-1}&\ldots &H_{t-1}&H_{t-1}&\ldots &H_{t-1}&H_{t-1}\\
\hline
\uzero&H_0&\ldots &\uzero&\uzero&\ldots &\uzero&\uzero\\
\uzero&H_{t-1}&\ldots &\ga_{1,\us_1-1}H_{t-1}&\ga_{1,\us_1}H_{t-1}&\ldots &\ga_{1,m-2}H_{t-1}&\ga_{1,m-1}H_{t-1}\\
\hline
\vdots &\vdots &\ddots &\vdots&\vdots&\ddots &\vdots&\vdots\\
\hline
\uzero&\uzero&\ldots &H_0&\uzero&\ldots &\uzero&\uzero\\
\uzero&\uzero&\ldots &H_1&\ga_{\us_1-1,\us_1}H_1&\ldots &\ga_{\us_1-1,m-2}H_1&\ga_{\us_1-1,m-1}H_1\\
\hline
\uzero&\uzero&\ldots &\uzero&H_0&\ldots &\uzero&\uzero\\
\hline
\vdots &\vdots &\ddots &\vdots&\vdots&\ddots &\vdots&\vdots\\
\hline
\uzero&\uzero&\ldots &\uzero&\uzero&\ldots &H_0&\uzero\\
\hline
\uzero&\uzero&\ldots &\uzero&\uzero&\ldots &\uzero&H_0\\
\end{array}
\right)
\end{eqnarray}

Using the pseudo upper-triangular parity-check matrix $\overset{\tiny\tr}{\bf
H}_{\sigma}(n;\uu)$ given by~(\ref{eqmain4}), we can decode the
(permuted) received array
${\bf v}_{\si}$ by successive
decoding of individual RS codes. Notice that $H_0$ is the
parity-check matrix of a RS code that can correct up to $u_0$ erasures,
and each $\left(H_0\atop H_i\right)$, $1\leq i\leq t-1$, is the
parity-check matrix of a RS code that can correct up to $u_i$
erasures. 

Once $\overset{\tiny\tr}{\bf H}_{\sigma}(n;\uu)$ has been obtained, 
the first step in the decoding algorithm is computing the $\sum_{i=0}^{t-1}u_is_i$
syndromes of ${\bf v}_{\si}$ (the permuted version of the received array
${\bf v}$) with respect to $\overset{\tiny\tr}{\bf H}_{\sigma}(n;\uu)$
(erasures are assumed to be zero in syndrome computation).
Since the number of erasures of ${\bf v}_{\si}$ is in non-increasing
order, the up to $u_0$ erasures in the last row of ${\bf v}_{\si}$ are
corrected by using the parity-check matrix $H_0$. Once this has been
done, the remaining $(\sum_{i=0}^{t-1}u_is_i)-u_0$ syndromes are
updated using the corrected information. The process is repeated with
each of the last $s_0$ rows of ${\bf v}_{\si}$, which contain up to $u_0$
erasures each. Once finished with correction of the last $s_0$ rows,
the next row, containing up to $u_1$ erasures, is corrected using the
parity-check matrix $\left(H_0\atop H_1\right)$. The process
continues by induction, until the first row, which contains up to
$u_{t-1}$ erasures, is corrected. Finally, the inverse permutation
$\si^{-1}$ is applied to the rows of the corrected version of
${\bf v}_{\si}$ to obtain the corrected version of ${\bf v}$.

Let us write formally the algorithm arising from the discussion
above. 

\begin{alg}[Decoding Algorithm]
\label{dec}
{\em
Consider a $t$-level GC-code $\C(n;\uu)$ as given by
Definition~\ref{defGC}. Let ${\bf v}$ be a received
$m\times n$ array with erasures. 

Let $\si :\{0,1,\ldots,m-1\}\;\ra\;\{0,1,\ldots,m-1\}$ be a permutation
of the rows of ${\bf v}$ and ${\bf v}_{\si}$ the array ${\bf v}$ with the rows
permuted according to $\si$, such that the number of erasures in each
row of ${\bf v}_{\si}$ is in non-increasing order. 

If the parity-check matrix ${\bf H} (n;\uu)$ of $\C(n;\uu)$ is
given by~(\ref{PCH}), consider the permuted parity-check matrix ${\bf
H}_{\si}(n;\uu)$ given by~(\ref{PCHp}), or, more in detail,
by~(\ref{eqmain3}), which corresponds to a 
permuted code $\C_{\sigma}(n;\uu)$.  

Let $\si^{-1}(j)\eq i_j$ for $0\leq j\leq m-1$.
Take the rectangular Vandermonde matrix given by~(\ref{eqmain5}) and,
by row operations, transform it into the upper triangular form given
by~(\ref{triang}). Use this upper triangular matrix to transform the
parity-check matrix ${\bf H}_{\sigma}(n;\uu)$ as
given by~(\ref{eqmain3}) into the pseudo upper triangular
parity-check matrix $\overset{\tiny\tr}{\bf H}_{\sigma}(n;\uu)$ given
by~(\ref{eqmain4}). Then proceed as follows:

\begin{enumerate}

\item Compute the $\sum_{i=0}^{m-1}u_is_i$ syndromes of ${\bf v}_{\si}$
with respect to the parity-check matrix $\overset{\tiny\tr}{\bf
H}_{\sigma}(n;\uu)$.

\item Correct the erasures in the last row of ${\bf v}_{\si}$ using the
RS parity-check 
matrix $H_0$ and the last $u_0$ syndromes. Then the next to last row
of ${\bf v}_{\si}$ using the RS parity-check
matrix $H_0$ and the next to last $u_0$ syndromes, and so on until
correcting the last $s_0$ rows. If any of these last rows had more
than $u_0$ erasures, then declare an uncorrectable error.

\item Using the corrected locations and values in the last $s_0$ rows
of ${\bf v}_{\si}$, update the first $\sum_{i=1}^{m-1}u_is_i$ syndromes
of ${\bf v}_{\si}$ with respect to $\overset{\tiny\tr}{\bf
H}_{\sigma}(n;\uu)$. 


\item Next, consider the last of the first $\sum_{i=1}^{m-1}s_i$ rows of
${\bf v}_{\si}$. If there are more than $u_{1}$ erasures in such
row, declare an uncorrectable error. Otherwise, correct up to $u_1$
erasures in the last of
these $\sum_{i=1}^{m-1}s_i$ rows 
using the last $u_{1}$ of the $\sum_{i=1}^{m-1}u_is_i$ syndromes with
respect to the RS code whose parity-check matrix is given by $\left(
{H_0\atop H_{1}}\right)$. Update then the first
$\left(\sum_{i=1}^{m-1}u_is_i\right)-u_{1}$ syndromes. 

\item Repeat the process until the first row, which contains up to
$u_{m-1}$ erasures, is corrected using the first $u_{m-1}$ syndromes
with respect to the RS code whose parity-check matrix is given by $\left(
{H_0\atop H_{m-1}}\right)$.

\item Obtain the corrected array ${\bf v}$ by applying the permutation
$\si^{-1}$ to the rows of the corrected array ${\bf v}_{\si}$.

\end{enumerate}

}
\end{alg}

The next example illustrates the decoding algorithm.

\begin{ex}
\label{exdec1}
{\rm
Let $n\eq 5$ and $\uu\eq (1,2,2,4)$.
Take the code $\C(5;(1,2,2,4))$ over the finite field ${\rm GF}(8)$ given
by the primitive polynomial
$1+x+x^3$. According to~(\ref{eqmain}) and~(\ref{eqmain2}), ${\bf H}
(5;(1,2,2,4))$ is given by the $9\times 20$ matrix

\begin{eqnarray*}
\left(\begin{array}{c|c|c|c}
H_0&\uzero&\uzero&\uzero\\
\uzero &H_0&\uzero &\uzero\\
\uzero &\uzero&H_0&\uzero\\
\uzero &\uzero&\uzero&H_0\\
\hline
H_2&H_2&H_2&H_2\\
\hline
H_1&\al^{-1}H_1&\al^{-2}H_1&\al^{-3}H_1\\
H_1&\al^{-2}H_1&\al^{-4}H_1&\al^{-6}H_1\\
\end{array}\right),
\end{eqnarray*}
where

\begin{eqnarray*}
H_0&=
\left(\begin{array}{ccccc}
1&1&1&1&1\\
\end{array}\right),
\end{eqnarray*}

\begin{eqnarray*}
H_2&=
\left(\begin{array}{ccccc}
\al^4&\al^3&\al^2&\al&1\\
\al^8&\al^6&\al^4&\al^2&1\\
\al^{12}&\al^9&\al^6&\al^3&1\\
\end{array}\right).
\end{eqnarray*}
and $H_1$ corresponds to the first row of $H_2$. Notice that
$\al^8\eq \al$, $\al^9\eq\al^2$ and $\al^{12}\eq \al^5$ since
$\al^7\eq 1$.

The codewords in the code are $4\times 5$ arrays. Assume that the
following array has been received:

\begin{eqnarray*}
{\bf v}&=&
\begin{array}{|c|c|c|c|c|}
\hline
E&\al^3&1&E&0\\
\hline
\al^6&E&E&E&E\\
\hline
\al^6&E&\al^5&E&1\\
\hline
\al^4&0&\al&E&\al^3\\
\hline
\end{array}
\end{eqnarray*}
where $E$ denotes an erasure.
We can see that there are 2 erasures in the first row, 4 in the
second, 2 in the third and one in the fourth. If we take the
permutation $$\si :\{0,1,2,3\}\;\ra\;\{0,1,2,3\}$$ such that
$\si(0)\eq 1$, $\si(1)\eq 0$, $\si(2)\eq 2$ and $\si(3)\eq 3$, the
permuted array is given by

\begin{eqnarray*}
{\bf v}_{\si}&=&
\begin{array}{|c|c|c|c|c|}
\hline
\al^6&E&E&E&E\\
\hline
E&\al^3&1&E&0\\
\hline
\al^6&E&\al^5&E&1\\
\hline
\al^4&0&\al&E&\al^3\\
\hline
\end{array}.
\end{eqnarray*}
We can see that the number of erasures in ${\bf v}_{\si}$ appears now in
non-increasing order: the first row has 4 erasures, the next two have
two, and the last row has one erasure. The parity-check matrix
${\bf H}_{\si}(5;(1,2,2,4))$ corresponding to the permuted code
$\C_{\sigma}(5;(1,2,2,4))$ is given by

\begin{eqnarray*}
{\bf H}_{\si}(5;(1,2,2,4))&=&
\left(\begin{array}{c|c|c|c}
H_0&\uzero&\uzero&\uzero\\
\uzero &H_0&\uzero &\uzero\\
\uzero &\uzero&H_0&\uzero\\
\uzero &\uzero&\uzero&H_0\\
\hline
H_2&H_2&H_2&H_2\\
\hline
\al^{-1}H_1&H_1&\al^{-2}H_1&\al^{-3}H_1\\
\al^{-2}H_1&H_1&\al^{-4}H_1&\al^{-6}H_1\\
\end{array}\right)
\end{eqnarray*}
and the matrix ${\bf \hat{H}}_{\sigma}(m;\uu)$ given by~(\ref{eqmain5}) is in this example

\begin{eqnarray*}
{\bf \hat{H}}_{\sigma}(4;(1,2,2,4)) \;\;=\;\;
\left(\begin{array}{cccc}
1&1&1&1\\
\al^{-1}&1&\al^{-2}&\al^{-3}\\
\al^{-2}&1&\al^{-4}&\al^{-6}\\
\end{array}\right) &=&
\left(\begin{array}{cccc}
1&1&1&1\\
\al^{6}&1&\al^{5}&\al^{4}\\
\al^{5}&1&\al^{3}&\al\\
\end{array}\right).
\end{eqnarray*}

Triangulating this last matrix, for instance, by Gaussian
elimination, we obtain the matrix given by~(\ref{triang})

\begin{eqnarray*}
\left(\begin{array}{cccc}
1&1&1&1\\
0&1&\al^{6}&\al\\
0&0&1&\al^3\\
\end{array}\right).
\end{eqnarray*}

Now, applying this matrix to the last 5
rows of ${\bf H}_{\si}(5;(1,2,2,4))$, we obtain 
the matrix given by~(\ref{triangb})

\begin{eqnarray*}
\left(\begin{array}{c|c|c|c}
H_2&H_2&H_2&H_2\\
\uzero &H_1&\al^{6}H_1&\al H_1\\
\uzero&\uzero&H_1&\al^{3}H_1\\
\end{array}\right).
\end{eqnarray*}

This matrix combined with the first 4 rows of ${\bf
H}_{\si}(5;(1,2,2,4))$, after some rearrangement, gives
the matrix of~(\ref{eqmain4}) as follows:

\begin{eqnarray*}
\overset{\tiny\tr}{\bf H}_{\sigma}(5;(1,2,2,4)) &=&
\left(\begin{array}{c|c|c|c}
H_0&\uzero&\uzero&\uzero\\
H_2&H_2&H_2&H_2\\
\hline
\uzero &H_0&\uzero &\uzero\\
\uzero &H_1&\al^{6}H_1&\al H_1\\
\hline
\uzero &\uzero&H_0&\uzero\\
\uzero&\uzero&H_1&\al^{3}H_1\\
\hline
\uzero &\uzero&\uzero&H_0\\
\end{array}\right).
\end{eqnarray*}

Next we compute the 9 syndromes of ${\bf v}_{\si}$ with respect to $\overset{\tiny\tr}{\bf
H}_{\sigma}(5;(1,2,2,4))$. Explicitly, these 9 syndromes are

\begin{eqnarray*}
{\bf S}_0 &=&\al^6\\
{\bf S}_1 &=&\al^3\\
{\bf S}_2 &=&\al^2\\
{\bf S}_3 &=&\al^3\\
{\bf S}_4 &=&\al\\
{\bf S}_5 &=&1\\
{\bf S}_6 &=&\al^3\\
{\bf S}_7 &=&\al^6\\
{\bf S}_8 &=&\al^5\\
\end{eqnarray*}

The first step is decoding one erasure in the fourth row of
${\bf v}_{\si}$, which corresponds to coordinate 18 of ${\bf v}_{\si}$ when
written as a vector. Since there is only one erased coordinate, such
erased coordinate has to equal the syndrome ${\bf S}_8 \eq\al^5$.
Thus, the last row of ${\bf v}_{\si}$ becomes

\begin{eqnarray*}
\left(\begin{array}{ccccc}
\al^4&0&\al&\al^5&\al^3.\\
\end{array}\right)
\end{eqnarray*}

The next step is updating the first 8 syndromes. Notice that
${\bf S}_0$, ${\bf S}_4$ and ${\bf S}_6$ remain the same since coordinate
18 of the corresponding rows in $\overset{\tiny\tr}{\bf H}_{\sigma}(5;(1,2,2,4))$ are
zero. As for the rest, using $\overset{\tiny\tr}{\bf H}_{\sigma}(5;(1,2,2,4))$,
we have

\begin{eqnarray*}
\begin{array}{lllllll}
{\bf S}_1 &=&{\bf S}_1\xor (\al)(\al^5)&=&\al^3\xor \al^6&=& \al^4\\
{\bf S}_2 &=&{\bf S}_2\xor (\al^2)(\al^5)&=&\al^2\xor 1&=&\al^6\\
{\bf S}_3 &=&{\bf S}_3\xor (\al^3)(\al^5)&=&\al^3\xor\al&=&1\\
{\bf S}_5 &=&{\bf S}_5\xor (\al^2)(\al^5)&=&1\xor 1&=&0\\
{\bf S}_7 &=&{\bf S}_7\xor (\al^4)(\al^5)&=&\al^6\xor \al^2&=&1\\
\end{array}
\end{eqnarray*}

Next we have to decode the two erasures corresponding to the third
row of ${\bf v}_{\si}$ using the parity-check matrix $\left({H_0\atop
H_1}\right)$ and the two syndromes ${\bf S}_6$ and ${\bf S}_7$.
Specifically, since erasures have occurred in locations 1 and 3 of
the third row, we have to solve the following system of two linear
equations with two unknowns:

\begin{eqnarray*}
X\xor Y&=&{\bf S}_6\;\eq\; \al^3\\
\al^3X\xor \al Y&=&{\bf S}_7\;\eq\;1.\\
\end{eqnarray*}

Solving this system, for instance by triangulation, gives $X\eq
\al^5$ and $Y\eq\al^2$. Replacing in the third row of ${\bf v}_{\si}$ gives

\begin{eqnarray*}
\left(\begin{array}{ccccc}
\al^6&\al^5&\al^5&\al^2&1\\
\end{array}\right).
\end{eqnarray*}

Next we need to update the first 6 syndromes, but as before,
syndromes ${\bf S}_0$ and ${\bf S}_4$ do not need to be updated. The
corrected erased coordinates correspond to coordinates 11 and 13 of
${\bf v}_{\si}$ when regarded as a vector. Again using $\overset{\tiny\tr}{\bf H}_{\sigma}(5;(1,2,2,4))$,
we have

\begin{eqnarray*}
\begin{array}{lllllll}
{\bf S}_1 &=&{\bf S}_1\xor (\al^3)(\al^5)\xor (\al)(\al^2)&=&\al^4\xor \al\xor\al^3&=& \al^5\\
{\bf S}_2 &=&{\bf S}_2\xor (\al^6)(\al^5)\xor (\al^2)(\al^2)&=&\al^6\xor\al^4\xor\al^4&=&\al^6\\
{\bf S}_3 &=&{\bf S}_3\xor (\al^2)(\al^5)\xor (\al^3)(\al^2)&=&1\xor 1\xor \al^5&=&\al^5\\
{\bf S}_5 &=&{\bf S}_5\xor (\al^2)(\al^5)\xor (1)(\al^2)&=&0\xor 1\xor\al^2&=&\al^6\\
\end{array}
\end{eqnarray*}

Now we have to decode the two erasures corresponding to the second
row of ${\bf v}_{\si}$ using the parity-check matrix $\left({H_0\atop
H_1}\right)$ and the two syndromes ${\bf S}_4$ and ${\bf S}_5$.
Since erasures have occurred in locations 0 and 3 of
the second row, we have to solve the following system of two linear
equations with two unknowns:

\begin{eqnarray*}
X\xor Y&=&{\bf S}_4\;\eq\; \al\\
\al^4X\xor \al Y&=&{\bf S}_5\;\eq\;\al^6.\\
\end{eqnarray*}

Solving this system gives $X\eq
\al^5$ and $Y\eq\al^6$. Replacing in the second row of ${\bf v}_{\si}$ gives

\begin{eqnarray*}
\left(\begin{array}{ccccc}
\al^5&\al^3&1&\al^6&0\\
\end{array}\right).
\end{eqnarray*}

Next we need to update the first 4 syndromes, but
syndrome ${\bf S}_0$ does not need to be updated. The
corrected erased coordinates correspond to coordinates 5 and 8 of
${\bf v}_{\si}$ when regarded as a vector. Using $\overset{\tiny\tr}{\bf H}_{\sigma}(5;(1,2,2,4))$,
we have

\begin{eqnarray*}
\begin{array}{lllllll}
{\bf S}_1 &=&{\bf S}_1\xor (\al^4)(\al^5)\xor (\al)(\al^6)&=&\al^5\xor \al^2\xor 1&=& \al\\
{\bf S}_2 &=&{\bf S}_2\xor (\al)(\al^5)\xor (\al^2)(\al^6)&=&\al^6\xor\al^6\xor\al&=&\al\\
{\bf S}_3 &=&{\bf S}_3\xor (\al^5)(\al^5)\xor (\al^3)(\al^6)&=&\al^5\xor \al^3\xor \al^2&=&0\\
\end{array}
\end{eqnarray*}

Finally we have to decode the four erasures corresponding to the first
row of ${\bf v}_{\si}$ using the parity-check matrix $\left({H_0\atop
H_2}\right)$ and the four syndromes ${\bf S}_0$, ${\bf S}_1$, ${\bf
S}_2$ and ${\bf S}_3$.
Since erasures have occurred in locations 1, 2, 3 and 4 of
the first row, we have to solve the following system of four linear
equations with four unknowns:

\begin{eqnarray*}
X\xor Y \xor Z\xor W&=&{\bf S}_0\;\eq\; \al^6\\
\al^3 X\xor \al^2 Y\xor \al Z\xor W&=&{\bf S}_1\;\eq\;\al\\
\al^6 X\xor \al^4 Y\xor \al^2 Z\xor W&=&{\bf S}_2\;\eq\;\al\\
\al^9 X\xor \al^6 Y\xor \al^3 Z\xor W&=&{\bf S}_3\;\eq\;0\\
\end{eqnarray*}

Solving this system, we obtain $X\eq
0$, $Y\eq\al^3$, $Z\eq 1$ and $W\eq \al^5$. Replacing in the first row of ${\bf v}_{\si}$ gives

\begin{eqnarray*}
\left(\begin{array}{ccccc}
\al^6&0&\al^3&1&\al^5\\
\end{array}\right).
\end{eqnarray*}

Finally, we apply $\si^{-1}$ (which in this particular case coincides
with $\si$) to the rows of the decoded version of ${\bf v}_{\si}$ to
obtain the decoded version of ${\bf v}$, giving the decoded array

\begin{eqnarray*}
{\bf v}&=&
\begin{array}{|c|c|c|c|c|}
\hline
\al^5&\al^3&1&\al^6&0\\
\hline
\al^6&0&\al^3&1&\al^5\\
\hline
\al^6&\al^5&\al^5&\al^2&1\\
\hline
\al^4&0&\al&\al^5&\al^3\\
\hline
\end{array}
\end{eqnarray*}

It can be verified that the syndromes of this array with respect to
the parity-check matrix ${\bf H}(5;(1,2,2,4))$ are zero, otherwise an
uncorrectable error would be declared.

\qed
}
\end{ex}

Let us point out that the decoding algorithm can be adapted to
correct errors as well as erasures (or combinations of both), but in
this paper we concentrate on the erasure problem only.

\subsection{Encoding}
\label{encoding}
The encoding is a special case of the decoding, where the parities
correspond to erasures. We can place the parities wherever we want as
long as the erasure-correcting capability of the code is not exceeded. A
natural choice is to put the parities in non-increasing order with
respect to their number in the last entries of each row. For
example, if $\uu\eq (1,2,2,4)$ like in Example~\ref{exdec1}, the
parities may be placed as follows (assuming $n\eq 5$ as in the
example):

$$
\begin{array}{|c|c|c|c|c|}
\hline
D&P&P&P&P\\
\hline
D&D&D&P&P\\
\hline
D&D&D&P&P\\
\hline
D&D&D&D&P\\
\hline
\end{array},
$$
where $D$ denotes data and $P$ parity.
In this case, the permutation $\si$ is the identity.
Knowing a priori where the parities are allows for precomputing the
pseudo-triangular parity-check matrix $\overset{\tiny\tr}{\bf H}_{\sigma}(n;\uu)$
given by~(\ref{eqmain4}). Then the encoding follows the steps of the
decoding to compute the parities. Let us retake the case of
Example~\ref{exdec1} to illustrate the encoding.

\begin{ex}
\label{exdec2}
{\rm
Assume that we want to encode the following array in
$\C(5;(1,2,2,4))$ over the finite field ${\rm GF}(8)$, where the entries
denoted by $P$ are the parities and are considered as erasures.

\begin{eqnarray*}
{\bf v}&=&
\begin{array}{|c|c|c|c|c|}
\hline
\al^5&P&P&P&P\\
\hline
\al^6&0&\al^3&P&P\\
\hline
\al^6&\al^5&\al^5&P&P\\
\hline
\al^4&0&\al&\al^5&P\\
\hline
\end{array}
\end{eqnarray*}

Following the decoding algorithm, as in Example~\ref{exdec1}, we need
to find the pseudo-triangular parity-check matrix $\overset{\tiny\tr}{\bf
H}_{\sigma}(5;(1,2,2,4)$, where in this case $\si$ is the identity
(the number of erasures in each row are already in non-increasing
order). Thus, ${\bf H}_{\si}(5;(1,2,2,4))\eq {\bf H}(5;(1,2,2,4))$
and the matrix given by~(\ref{eqmain5}) is

\begin{eqnarray*}
{\bf \hat{H}}_{\sigma}(4;(1,2,2,4)) \;\;=\;\;
\left(\begin{array}{cccc}
1&1&1&1\\
1&\al^{-1}&\al^{-2}&\al^{-3}\\
1&\al^{-2}&\al^{-4}&\al^{-6}\\
\end{array}\right)
&=&
\left(\begin{array}{cccc}
1&1&1&1\\
1&\al^{6}&\al^{5}&\al^{4}\\
1&\al^{5}&\al^{3}&\al\\
\end{array}\right)
\end{eqnarray*}

Triangulating this last matrix, for instance, by Gaussian
elimination, we obtain the matrix of~(\ref{triang})

\begin{eqnarray*}
\left(\begin{array}{cccc}
1&1&1&1\\
0&1&\al^{2}&\al^3\\
0&0&1&\al^3\\
\end{array}\right).
\end{eqnarray*}

Now, with this matrix, we can obtain $\overset{\tiny\tr}{\bf
H}_{\sigma}(5;(1,2,2,4))$
given by~(\ref{eqmain4}) as follows:

\begin{eqnarray*}
\overset{\tiny\tr}{\bf H}_{\sigma}(5;(1,2,2,4))
&=&
\left(\begin{array}{c|c|c|c}
H_0&\uzero&\uzero&\uzero\\
H_2&H_2&H_2&H_2\\
\hline
\uzero &H_0&\uzero &\uzero\\
\uzero &H_1&\al^{2}H_1&\al^3 H_1\\
\hline
\uzero &\uzero&H_0&\uzero\\
\uzero&\uzero&H_1&\al^{3}H_1\\
\hline
\uzero &\uzero&\uzero&H_0\\
\end{array}\right).
\end{eqnarray*}

The encoding now proceeds like the decoding using this parity-check
matrix $\overset{\tiny\tr}{\bf H}_{\sigma}(5;(1,2,2,4))$. Doing so, it can be
verified that the encoded array coincides with the decoded array of
Example~\ref{exdec1}. Since $\overset{\tiny\tr}{\bf H}_{\sigma}(5;(1,2,2,4))$ is
precomputed, the encoding starts at this point, saving the time
necessary to produce this matrix, as in the general decoding
algorithm.

\qed
}
\end{ex}

\section{Conclusions}
We have presented a method of implementing Generalized Concatenated
Codes as erasure-correcting codes over $m\times n$ arrays. We
proved the fundamental properties of the codes and gave efficient
encoding and decoding algorithms. In general, GC codes are weaker
than PMDS codes, but the trade-off is that they allow for a smaller
field, mainly, the size of the field is given by the length of the
rows of the arrays.

\end{document}